\documentclass[notitlepage,nofootinbib,preprintnumbers,amssymb,superscriptaddress]{revtex4-1}
\usepackage{amssymb,float,amsmath,amsfonts,amsthm,cases,bm,latexsym,mathrsfs, eufrak, graphicx, color, epsfig, wrapfig, comment, xcolor, titlesec}
\usepackage[normalem]{ulem}
\usepackage{cancel}
\definecolor{linkcolor}{HTML}{799B03}
\definecolor{urlcolor}{HTML}{799B03}
\definecolor{ultramarine}{rgb}{0.07, 0.04, 0.56}
\definecolor{cadmiumgreen}{rgb}{0.0, 0.42, 0.24}
\definecolor{indigo(dye)}{rgb}{0.0, 0.25, 0.42}
\usepackage[linktocpage=true]{hyperref}
\hypersetup{
colorlinks=true,
citecolor=green,
linkcolor=cadmiumgreen,
urlcolor=indigo(dye),
}

\usepackage{autobreak}

\def\[{\begin{equation}}
\def\]{\end{equation}}

\begin{document}
\begin{center}

\hfill ITEP/TH-32/25\\

\end{center}

\vspace{1cm}
\title{Towards the $\tau$-function of the quantum groups}

\author{M. Chepurnoi}
\email{chepurnoi.ma22@physics.msu.ru}
\affiliation{NRC, "Kurchatov Institute", 123182, Moscow, Russia}
\affiliation{Institute for Theoretical and Mathematical Physics,
MSU, 119991 Moscow, Russia}
\author{M. Sharov}
\email{sharov.mr22@physics.msu.ru}
\affiliation{NRC, "Kurchatov Institute", 123182, Moscow, Russia}
\affiliation{Institute for Theoretical and Mathematical Physics,
MSU, 119991 Moscow, Russia}
\affiliation{Institute for Nuclear Research of the Russian Academy of Sciences,
60th October Anniversary Prospect, 7a, 117312 Moscow, Russia}

\begin{abstract}
    Non-perturbative partition functions of quantum theories constitute a class of $\tau-$functions, which are distinguished satisfying Hirota's bilinear identities(BI). To make this statement general, there must be a proper definition of $\tau-$function that gives rise to a set of bilinear identities. In the classical definition of $\tau-$function for integrable Toda or KP hierarchies, there is a restriction on matrix elements to be based on group-like elements with the comultiplication $\Delta(g)=g \otimes g$. This restriction can not be straightforwardly transferred to the q-deformed case, because there are no group-like elements in q-deformed universal enveloping algebra (UEA), except for its Cartan subalgebra. The new approach to the $\tau-$function is to remove the restriction on g to be obligatory the group-like element. The main result of this work is a derivation of the set of bilinear identities and $\tau-$functions for $U_q(\mathfrak{sl}_3)$ in the fundamental representations for non-group-like elements. We consider difference operators which lead to the basic bilinear identities. Also, we provide an analysis of the ways of obtaining BI for higher rank algebras $U_q(sl_n)$.
\end{abstract}

\maketitle

\section{Introduction}
The two-dimensional Toda hierarchy is an integrable system of difference-differential equations, originally developed in \cite{2TDL} as a generalization of the KP hierarchy. The well-known 2D Toda lattice(2TDL) equation—serves as the foundation of the hierarchy. Using the results obtained by the Kyoto School \cite{DJKM1,DJKM}, a family of solutions can be derived through the fermionic representation of the $\mathfrak{gl}(\infty)$ algebra. These solutions are parameterized by elements of the group $GL(\infty)$.


The whole set of Hirota-like equations can be solved by a single function called $\tau$-function \cite{GKLMM,gentau,2312,KMM}, which, in a broad sense, can be defined as a generating function for matrix elements:
\begin{equation}
\label{gtau}
\tau_V(t,\bar t|g) \equiv
\sum_{m,\bar m\in V}
s^V_{m,\bar m}(t,\bar t)
\left.\langle m | g | \bar m
\rangle\right._V,
\end{equation}
where $V$ is a Verma module, and $g$ is the group-like elements, i.e. such elements that  its comultiplication satisfies $\Delta(g)=g \otimes g$. Solution of integrable hierarchies are parametrized by points on the Universal Grassmannian, which are labeled by such g. For simple Lie algebras, it is possible to construct group-like elements with based a corresponding comultiplication. For group-like elements, the most conventional way of obtaining bilinear identities is the intertwining operator method. The intertwiners become fermions within the context of 2TDL. Some examples of this method were described in detail in \cite{GKLMM,gentau}, especially for the fundamental representations of $U(sl_n)$, which corresponds to the 2d Toda system. The requirement of the fundamental representation is crucial for deriving the bilinear identities on the $\tau$-function, and currently there are no known ways to eliminate this restriction.


In the context of q-deformed algebras, there are no group-like elements, except for the Cartan subalgebra, because the comultiplication is deformed. A possible method to construct these elements involves choosing $g$ such that $g \in U(\mathcal{G}) \otimes A$, where $A$ is the (non-commutative) algebra of functions on the quantum group \cite{MV}. The bilinear identity and the $\tau$-function for fundamental representations of $U_q(\mathfrak{sl}_n)$ were already obtained \cite{KMM} with the use of q-deformed intertwining operators (q-fermions). However, in the q-deformed case, the classical meaning of integrability as existing commutative flows was lost. In the non-deformed case, various parametrizations of the $\tau$-function exist. Among these, one of the parametrizations is associated with commutative flows, but its analogue for q-deformed case is not yet constructed has not yet been subjected to q-deformation.

Another way \cite{2312, Bourg} to obtain the bilinear identities for the q-deformed algebras is to remove the requirement of the group-like element and consider $\tau$-functions based on an arbitrary element $X$ in q-UEA with 
\begin{equation}
\Delta(X)=\sum\limits_\alpha X_\alpha'\otimes X_\alpha'' \;,
\end{equation}
where $X,\, X_\alpha',\, X_\alpha''\, \in\, U_q(\mathcal{G})$.

The paper is organized as follows: firstly, we discuss the case of the $\tau$-function based on the group-like element and show the way of derivation of the BI. Then we consider the scenarios of arbitrary UEA elements in the $\tau$-function and perform the derivation of the BI for $U_q(sl_3)$. Finally, we analyze the intertwining operators for $U_q(\mathfrak {sl} _ n)$ and investigate the possibility of constructing the BI based on q-intertwiners (q-fermions) for non-group-like elements.

\section{Group-like element}

\subsection{Intertwining operator}

Construction of intertwining operators as a tool for of bilinear identities is a generalization of classical KP/Toda hierarchies. This method is presented for finite-dimensional Lie algebras and their q-deformed counterparts. V is a Verma module. W is some irreducible finite-dimensional representation of UEA. Bilinear identities are derived for four Verma modules $V,\hat{V},V',\hat{V}'$.

1. Starting point is embedding of Verma modules $\hat{V}$, $\hat{V'}$ in tensor product $V \otimes W$ and $W'\otimes V'$ respectively. Consider, we define two respective embedding operators: right and left
\begin{subequations}
\begin{align}
   & E_R:\hat{V} \longrightarrow V \otimes W ,\\ &
    E_L':\hat{V'}\longrightarrow W'\otimes V'.
\end{align} 
\end{subequations}

Intertwining operators are characterized by satisfying the following property:
\begin{equation} \label{intproperty}
 a \Phi=\Phi\Delta(a)  
\end{equation}

Fixing representations V and W removes 
arbitrariness in choosing $\hat{V}$: only a finite number of possible compatible representations remain.
These two operators are a generalization of fermions.

2. The next step is projection. We assume that the product $W \otimes W'$ contains a unit representation of UEA. We define the projector operator as follows
\begin{equation}
    \pi: \ \ W\otimes W' \longrightarrow I.
\end{equation}

Using this operator, it is possible to build a new intertwining operator:
\begin{equation}
\label{Gamma}
\Gamma: \ \
\widehat V \otimes \widehat V' \stackrel{E_R\otimes E_L'}{\longrightarrow}
V \otimes W \otimes W' \otimes V'
\stackrel{I\otimes \pi \otimes I}{\longrightarrow} V \otimes V'.
\end{equation}

As the consequence of \eqref{intproperty}, this operator commutes with the group-like element's coproduct:
\begin{equation}
    \Gamma( g\otimes g)=(g \otimes g) \Gamma.
\end{equation}

We call this relation a basic bilinear relation. Taking matrix elements allows one to obtain the bilinear identity:
\begin{equation}
\begin{aligned}
&\Big\langle n\Big|U(t)\otimes \Big\langle m\Big|U(t')\Gamma\Delta(g)\bar U(\bar t)\Big|n-1\Big\rangle\otimes \bar U (\bar t')\Big|m+1\Big\rangle= \\ &
=\Big\langle n\Big|U(t)\otimes \Big\langle m\Big|U(t')\Delta(g)\Gamma \bar U(\bar t)\Big|n-1\Big\rangle\otimes\bar U(\bar t')\Big|m+1\Big\rangle,
\end{aligned}
\end{equation}

where $U(t),U(t'),\bar U(\bar t),\bar U(\bar t')$ are the evolution operators, which we define later. And then it can be possible to rewrite bilinear identities in terms of the $\tau-$function, using differential operators along the lines of \cite{DJKM1, DJKM}.

\subsection{Group-like element for q-UEA.}
In the quantum case, parameters, labeling the “group manifold,” are no longer commuting c-numbers \cite{MV}. In the q-deformed UEA, the general construction of the group-like element, which possesses the property $\Delta(g)=g\otimes g$, cannot lead to the group-like elements lying in the UEA. This problem appears because such elements exist only in the Cartan subalgebra of q-UEA due to the deformed comultiplication of algebra generators, which in a general sense reads as 
\begin{equation}
    \Delta(T^{(\alpha)})\equiv D^{\alpha}_{\beta\gamma}T^{(\beta)}\otimes
T^{(\gamma)},
\end{equation}
but
\begin{equation}
    \Delta(K_{(\alpha)}) = K_{(\alpha)} \otimes K_{(\alpha)}.
\end{equation}
We choose the group-like element (also known as “universal T-matrix”) in the following way
\begin{equation}\label{grouel}
\hbox{{\bf T}}\equiv\sum_{\alpha}X^{(\alpha)}\otimes T^{(\alpha)}\in A(G)
\otimes U_q({\cal G}).
\end{equation}

Such T should still satisfy group-like element comultiplication, which reads
\begin{equation} \label{qcom}
  \Delta_U(\hbox{{\bf T}})=\hbox{{\bf T}}\otimes_U\hbox{{\bf T}} \in A(G)\otimes U_q({\cal G})\otimes U_q({\cal G})
\end{equation}

In this context, for simple Lie algebras, the group-like element appears as
\begin{equation}
\begin{aligned}
g &= g_Ug_Dg_L, \\
g_U =\left.\prod_s\right.^<\ {e}_{q^{-1}}^{\theta_se_{i(s)}},   \ \ \
g_D = &\prod_{i=1}^{r_{G}} e^{\vec\phi\vec h},   \ \ \
g_L = \left.\prod_s\right.^>\ {e}_q^{\chi_sf_{i(s)}}.
\end{aligned}
\end{equation}

Where $e_q(x)$ is the q-exponential:
\begin{equation}
e_q(x)=\sum_{n \geq  0} \frac{x^n}{[n]_q!}q^{\frac{n(n-1)}{2}}.
\end{equation}

In thus parametrization it is crucial, that only the simple roots appear, because it leads to the Heisenberg-like realization of the algebra of functions $\chi$, $\phi$, $\theta$, and this is advantageous from the technical point of view. Simple roots can appear multiple times in the product to span the entire Borel subalgebra. The explicit form of the realization reads as
\begin{equation} \label{par A}
\{T_A\} = \{\prod_sT^{n(s)}_{i(s)} \}.
\end{equation}


Note that upon introducing the algebra of functions, a non-trivial $\mathcal{R}$-matrix appears within the framework. For the group-like element $\hat{g} \in \mathcal{G}\otimes \mathcal{A}$, $\mathcal{R}$-matrix, defined as follows:
\begin{equation}
    (I\otimes_U \hat{g}) \cdot (\hat{g} \otimes_U I) = \mathcal{R} \cdot (\hat{g} \otimes_U I) \cdot (I \otimes_U \hat{g}) \cdot \mathcal{R}^{-1} = \mathcal{R} \cdot \Delta_U g  \cdot \mathcal{R}^{-1}.
\end{equation}

Nontrivial $\cal R$-matrix appears due to non-commutative $\chi$, $\phi$, $\psi$. With the selected parametrization of $g$, it can be derived \cite{MV} that $\cal R$ \textbf{is definitely the well-known $\cal R$-matrix}, defined for algebra $U_q(sl_n)$, $\mathcal{R}: \; V_1 \otimes V_2 \rightarrow V_2 \otimes V_1$. For the all simple roots
\begin{equation}
\begin{aligned}
\Delta\left(e_{i}\right) =I \otimes e_{i}&+e_{+i} \otimes q_i^{-2 h_i}=\mathcal{R}^{-1}\left(q_i^{-2h _i} \otimes e_{i}+e_{i} \otimes I\right) \mathcal{R},\\
\mathcal{R}\left(q^{2 \vec{\phi} \vec{h}} \otimes q^{2 \vec{\phi} \vec{h}}\right)&=\left(q^{2 \vec{\phi} \vec{h}} \otimes q^{2 \vec{\phi} \vec{h}}\right) \mathcal{R}.
\end{aligned}
\end{equation}

Then, the $\mathcal{R}$-matrix has the explicit form
\begin{equation} \label{Rmat}
\mathcal{R} = q^{-\sum_{ij}^{r_G} ||\vec{\alpha}_i||^2 ||\vec{\alpha}_j||^2 (\vec{\alpha}\vec{\alpha})^{-1}_{ij} h_i \otimes h_j} 
\prod_{\vec{\alpha} > 0}^{\mathit{d_B}} \mathcal{E}_{q_{\vec{\alpha}}} 
\left( - (q_{\vec{\alpha}} - q_{\vec{\alpha}}^{-1}) e_{\vec{\alpha}} \otimes f_{\vec{\alpha}} \right),
\end{equation}
where we define $q_{\vec{\alpha}}$ as $q_{\vec{\alpha}}=q^\frac{{{||\vec{\alpha}||}^2}}{2}$
\section{Beyond the group-like element}
\subsection{Split Casimir}

In the present paper, we work with a non-group-like element $X \in U(\mathcal{G})$, and it is possible to consider different basic bilinear relations. The Split Casimir or any central element can be used instead of intertwining operators. The essence of the method lies in the following property of the split Casimir operator \cite{Isaev}:
\begin{equation} \label{BBI1}
\Delta(X)\mathcal{C}=\mathcal{C}\Delta(X)
\end{equation}

Here, we use that the center of the classical algebra is preserved under q-deformation \cite{qIsaev, FRT}, this statement is not trivial, because an arbitrary deformation of the commutation relations may make the center trivial.

Taking the matrix element of \eqref{BBI1} yields a bilinear identity, which is generally not the identity on the $\tau$-function.

\begin{equation}
\begin{aligned} \label{BI_C}
&\Big\langle n\Big|U(t)\otimes \Big\langle m\Big|U(t')\mathcal{C}\Delta(X)\bar U(\bar t)\Big|n\Big\rangle\otimes \bar U (\bar t')\Big|m\Big\rangle= \\ &
=\Big\langle n\Big|U(t)\otimes \Big\langle m\Big|U(t')\Delta(X)\mathcal{C}  \bar U(\bar t)\Big|n\Big\rangle\otimes\bar U(\bar t')\Big|m\Big\rangle.
\end{aligned}
\end{equation}
Pushing Casimir to the left vectors through the evolution operators at the l.h.s. of the equation, and to the right vectors at the r.h.s., we obtain a bilinear identity and study the possibility of a recovery of $\tau$-function is possible.

\section{Bilinear identity for $U_q(\mathfrak{sl_3})$}
\subsection{Algebra of $U_q(\mathfrak{sl}_n)$}
Recall the commutation relations for $U_q(\mathfrak{sl}_3)$(we follow \cite{2312} in our notations):
\begin{subequations}
\begin{align}
[e_i,f_j]=\delta_{ij}\frac{q^{h_i}-q^{-h_i}}{q-q^{-1}},\\
q^{h_i}e_j q^{-h_i}=q^{a_{ij}}e_j,\\
q^{h_i}f_j q^{-h_i}=q^{-a_{ij}}f_j,\\
q^{h_i}q^{h_j}=q^{h_j}q^{h_i},\\
\end{align} 
\end{subequations}

here $a_{ij}$ is Cartan matrix. Comultiplication is given by:
\begin{subequations}
\begin{align} 
\Delta(e_i)=q^\frac{h_i}{2}\otimes e_i + e_i  \otimes q^{-\frac{h_i}{2}},\\
\Delta(f_i)=q^\frac{h_i}{2}\otimes f_i + f_i  \otimes q^{-\frac{h_i}{2}},\\
\Delta({q^{h_i}})=q^{h_i} \otimes q^{h_i}.
\end{align} 
\end{subequations}

\subsection{Evolution operators.} \label{sec:fund_repr}

In what follows, we work only with fundamental representations of $U_q(\mathfrak{sl_3})$ because they are only ones where the bilinear identity may be written down in terms of the $\tau$-function.

We choose evolution operators as follows:
\begin{equation} \label{evol_op_U}
    U(t)=\exp_q \left(\sum_{k=1}^{N-1} t_k \mathbf{E}_k^{(n)}\right), \quad
    \bar U(\bar t)=\exp_{q^{-1}} \left(\sum_{k=1}^{N-1} \bar{t}_k \mathbf{F}_k^{(n)}\right),
\end{equation}

Indices n correspond to the n-th fundamental representation. Where $\mathbf{E}^{(n)}_k$ and $\mathbf{F}_k^{(n)}$ and defined as follows

\begin{equation} \label{E_bold}
\mathbf{E}_k^{(n)}:=\left(\sum_i {e}_i\right)^k, \quad \mathbf{F}_k^{(n)}:=\left(\sum_i {f}_i\right)^k,
\end{equation}

Where indices i correspond to the i-th simple root. From \eqref{E_bold} follows that the set of flows $\bf{E}^{(n)}_k$ mutually commute, which is also true for flows $\bf{F}^{(n)}_k$.

Operators $\bf{E}^{(n)}_k$, $\bf{F}^{(n)}_k$, which we choose as flows in evolution operators, generate all space of fundamental representation of UEA \cite{gentau}. This statement becomes clear in the explicit matrix form of the fundamental representation.  $\bf{E}^{(n)}_k$ and $\mathbf{F}_k^{(n)}$ are the sets of commutative operators, which correspond to the classical meaning of integrability as commutative flows.
 
For these operators in the $U_q(\mathfrak{sl}_3)$ $\bf{E}^{(n)}_1$, $\bf{E}^{(n)}_2$, and $\mathbf{F}_1^{(n)}$, $\mathbf{F}_2^{(n)}$ the following property is accomplished:
\begin{equation}\label{zeroeq}
\mathbf{E}_1^{(n)} \mathbf{E}_2^{(n)} = 0, \quad \mathbf{{F}^{(n)}_1}\mathbf{{F}^{(n)}_2}=0
\end{equation}

It is correct only in the case of $U_q(\mathfrak{sl}_3)$. The generalization of this statement to algebras of higher ranks does not exist.
$\tau$-function, which corresponds to the n-th fundamental representation:
\begin{equation}
\tau_n(t ; X)={ }_n\langle 0| \exp_{q} \left(\sum_{k=1}^{N-1} t_k \mathbf{E}_k^{(n)}\right) X \exp_{q^{-1}} \left(\sum_{k=1}^{N-1} \bar{t}_k \mathbf{F}_k^{(n)}\right)|0\rangle_n.
\end{equation}

\subsection{Bilinear identity.} \label{Sec:BI}
As we mentioned before, we use the split Casimir approach. The expression for the quadratic Casimir in fundamental representations of $U_q(\mathfrak{sl}_3)$ has explicit form \cite{Casimir}
\begin{equation}    
\begin{aligned} 
\mathcal{C}_{2 q}= & \frac{1}{2}\left(q^2+q^{-2}\right) \\ & \times\left(\left[\frac{2 H_1}{3}+\frac{H_2}{3}\right]_q^2+\left[\frac{H_1}{3}+\frac{2 H_2}{3}\right]_q^2\right)+\left[\frac{H_1}{3}-\frac{H_2}{3}\right]_q^2 \\ & +\frac{1}{2}\left(K K_2 q^{-1}+K^{-1} K_2^{-1} q\right) X_1^{+} X_1^{-}+\frac{1}{2}\left(K K_2 q+K^{-1} K_2^{-1} q^{-1}\right) X_1^{-} X_1^{+} \\ & +\frac{1}{2}\left(K K_1^{-1} q+K^{-1} K_1 q^{-1}\right) X_2^{+} X_2^{-}+\frac{1}{2}\left(K K_1^{-1} q^{-1}+K^{-1} K_1 q\right) X_2^{-} X_2^{+} \\ & +\frac{1}{2} K\left(X_3^{+} \widetilde{X}_3^{-}+\widetilde{X}_3^{-} X_3^{+}\right)+\frac{1}{2} K^{-1}\left(\widetilde{X}_3^{+} X_3^{-}+X_3^{-} \widetilde{X}_3^{+}\right),
\end{aligned}
\end{equation}
where $K=K_1^{\frac{1}{3}}K_2^{-\frac{1}{3}}$.

But for simplifying calculations, since the commutative property preserves, we may use another element of the center instead of the Casimir operator \cite{Casimir}
\begin{equation}
\begin{aligned}
Z_1= & K\left[K_2 X_1^{+} X_1^{-} q^{-1}+X_3^{+} \tilde{X}_3^{-}+K_1^{-1} X_2^{+} X_2^{-} q\right. \\
& +\left(K_1 K_2 q^{-2}+K_1^{-1} K_2+K_1^{-1} K_2^{-1} q^2\right) \\
& \left.\times\left(q-q^{-1}\right)^{-2}\right]
\end{aligned}
\end{equation}

$X_3$ corresponds to the third root, which is defined as follows
\begin{subequations}
\begin{align} 
X_3^{ \pm} \equiv q^{1 / 2} X_1^{ \pm} X_2^{ \pm}-q^{-1 / 2} X_2^{ \pm} X_1^{ \pm},\\
\widetilde{X}_3^{ \pm} \equiv-q^{-1 / 2} X_1^{ \pm} X_2^{ \pm}+q^{1 / 2} X_2^{ \pm} X_1^{ \pm}.
\end{align}
\end{subequations}

Taking the matrix element as was described above \eqref{BI_C}, we obtain the bilinear identity
\begin{equation}
\begin{aligned}\label{bi}
    &{}_{\lambda'}^{}\langle0|e_q^{\sum_{k=1}^{N-1} t_k '\mathbf{E}_k^{(n)}} {}_\lambda\langle0|e_q^{\sum_{k=1}^{N-1} t_k \mathbf{E}_k^{(n)}} \Delta(X) \Delta \left(Z_1\right) e_{q^{-1}}^{\sum_{k=1}^{N-1} t_k \mathbf{F}_k^{(n)}}|0\rangle_{\lambda}e_{q^{-1}}^{\sum_{k=1}^{N-1} t_k' \mathbf{F}_k^{(n)}}|0\rangle_{\lambda'}=\\ & {}_{\lambda'}^{}\langle0| e_q^{\sum_{k=1}^{N-1} t_k '\mathbf{E}_k^{(n)}} {}_\lambda\langle0|e_q^{\sum_{k=1}^{N-1} t_k \mathbf{E}_k^{(n)}} \Delta \left(Z_1\right) \Delta(X)e_{q^{-1}}^{\sum_{k=1}^{N-1} t_k \mathbf{F}_k^{(n)}}|0\rangle_{\lambda}e_{q^{-1}}^{\sum_{k=1}^{N-1} t_k' \mathbf{F}_k^{(n)}}|0\rangle_{\lambda'}.
\end{aligned}
\end{equation}

To obtain the bilinear identity in difference form, we use the q-deformed differential operator 
\begin{equation}\label{qdiff}
    D_{q,x}f(x)=\frac{f(q^2x)-f(x)}{(q^2-1)x}.
\end{equation}

This operator acts on the q-exponent as follows
\begin{equation}\label{diffexp}
     D_{q,x}e_q^{xT}=T e_q^{xT},
\end{equation}

And obey the Leibniz rule:
\begin{equation}
    D_q(f(x)g(x))=f(x)D_q(g(x))+g(q^2x)D_q(f(x)).
\end{equation}

Bringing each part of tensor product of the central element $\Delta (Z_1)$ through the evolution operators is a separate task. For this purpose, we use the q-deformed Baker-Campbell-Hausdorff formula \cite{BCH,MV}
\begin{equation}
    e_{q}^{A}Be_{q^{-1}}^{-A}=B+[A,B]+\frac{1}{[2]_q!}[A,[A,B]_q]+\frac{1}{[3]_q!}[A,[A,[A.B]_{q}]_{q^2}]+...
\end{equation}

The q-deformed commutator is defined as follows:
\begin{equation} \label{qcom}
    [A,B]_{q^k}=q^kAB-q^{-k}BA.
\end{equation}

 Action on vectors yields the bilinear identity. The final step is to rewrite the bilinear identity in terms of the $\tau$-function. 
 
 For q-commutative operators the following property is carried out \cite{klim,qcalc}
\begin{equation} \label{split_exp}
    e_q^{t_1 A+ t_2B}=e_q^{t_1{A}}e_q^{t_2 B} , \quad \text{if} \quad [A,B]_{q^{\frac{1}{2}}}= 0.
\end{equation}
 
Here we use the fact that flows $\bf{E}^{(n)}_k$, $\bf{F}^{(n)}_k$ possess the property \eqref{zeroeq}, which formally makes them to be q-commutative(hold on only for $U_q(\mathfrak{sl}_3)$). Subsequently, employing \eqref{diffexp} results in the bilinear identity on the $\tau$-function

\begin{equation}
\begin{aligned}
   &\frac {q^\frac
  {2}{3}}{(1 + q^2)^2}  \Biggr[-(1 + q^2) ((1 + q^2) (-\bar{t_1'} + 
            q \bar{t_1}) (-1 + (\bar{t_1'} q - \bar{t_1}) D_{\bar{t_1}} + (q (\bar{t_1'}^3 - 
               \bar{t_1'} \bar{t_2'} (1 + q^2) + (\bar{t_2'} + \bar{t_2'} q^2 - \bar{t_1'}^2 (2 + q^2)) \bar{t_1} +\\&
               \bar{t_1'} q (2 + q^2) \bar{t_1}^2 - q^2 \bar{t_1}^3) + (1 + q^2) (\bar{t_1'} - 
               q \bar{t_1}) \bar{t_2}) D_{\bar{t_2}}))D_{\bar{t_1'}}+ ((1 + q^2) (-\bar{t_2'} (1 + q^2) + 
          q (-\bar{t_1'}^2 q + \bar{t_1'} (1 + q^2) \bar{t_1} - \bar{t_1}^2 + \bar{t_2} + q^2 \bar{t_2})) -\\& (1 + 
          q^2) (-\bar{t_1'}^3 q^3 + \bar{t_1'}^2 (1 + q^2 + q^4) \bar{t_1} + 
          \bar{t_1'} q (-\bar{t_2'} (1 + q^2) - (2 + q^2) \bar{t_1}^2 + \bar{t_2} + q^2 \bar{t_2}) + 
          q \bar{t_1} (\bar{t_2'} (q + q^3) + \bar{t_1}^2 - (1 + q^2) \bar{t_2})) D_{\bar{t_1}} + \\&(-\bar{t_1'}^4 q^3 + \bar{t_1'}^3 q (1 + q^2)^2 \bar{t_1} + 
          \bar{t_1'} q (1 + q^2)^2 \bar{t_1}^3 + 
          \bar{t_1'}^2 (\bar{t_2'} q (-1 + q^4) - (1 + 2 q^2 + 2 q^4 + 
                q^6) \bar{t_1}^2) + (\bar{t_2'} (q + q^3) + 
             \bar{t_1}^2 - \\&(1 + q^2) \bar{t_2}) (\bar{t_2'} (1 + q^2) - 
             q (\bar{t_2} + q^2 (\bar{t_1}^2 + \bar{t_2})))) D_{\bar{t_2}})D_{\bar{t_2'}} + (1 + 
       q^2) (\bar{t_1'} q D_{\bar{t_1}} + \bar{t_1'} q^3 D_{\bar{t_1}} - 
       q^2 \bar{t_1} D_{\bar{t_1}} - q^4 \bar{t_1} D_{\bar{t_1}} - 
       \bar{t_1'}^2 q D_{\bar{t_2}} +\\& \bar{t_2'} q D_{\bar{t_2}} + 
       \bar{t_2'} q^3 D_{\bar{t_2}} + \bar{t_1'} q \bar{t_1} D_{\bar{t_2}} + 
       \bar{t_1'} q^3 \bar{t_1} D_{\bar{t_2}} - q^4 \bar{t_1}^2 D_{\bar{t_2}} - 
       q^2 \bar{t_2} D_{\bar{t_2}} - q^4 \bar{t_2} D_{\bar{t_2}} + 
       q t_1 D_{t_1'} + \\&
       q^3 t_1 D_{t_1'} - t_1' D_{t_1'} - 
       q^2 t_1' D_{t_1'} + (t_1'^2 + q (t_2 - t_1 t_1') + 
          q^3 (t_1^2 + t_2 - t_1 t_1') - t_2' - 
          q^2 t_2') D_{t_2'} + \\
       &((1 + q^2) (q t_1 - t_1') (-q + (-t_1 + q t_1') D_{t_1'} + (-t_1 t_1'^2 + q^4 t_1 t_2' + 
             q^2 t_1 (-2 t_1'^2 + t_2') + 
             q (-t_1 t_2 + t_1^2 t_1' + 
                t_2 t_1' + t_1'^3 - t_1' t_2') - \\&
             q^3 (t_1^3 + t_1 t_2 - 2 t_1^2 t_1' - 
                t_2 t_1' + t_1' t_2')) D_{t_2'}) + 
        (q (q (t_1^2 - t_2) - q^3 t_2 - t_1 t_1' + t_2' + 
             q^2 (-t_1 t_1' + t_1'^2 + t_2')) - (q^4 t_1^2 
t_1' + (t_1^2 - t_2) t_1' +\\& q^2 (t_1^2 - t_2) t_1' +
             q^3 (t_1 (t_2 - 
                   2 t_1'^2 - t_2') + t_1' (t_1'^2 + 
t_2')) - 
             q (t_1^3 - t_1' t_2' + 
                t_1 (-t_2 + t_1'^2 + t_2'))) D_{t_1'} + ((t_1^2 - t_2) t_1'^2 + t_2 t_2' +\\& 
             q^4 ((t_1^2 + t_2) t_1'^2 + t_2 t_2') +
             q^2 (t_1^2 t_1'^2 + 2 t_2 t_2') - 
             q (-t_1^2 t_2 + t_2^2 + t_1^3 t_1' + 
                t_1 t_1'^3 - t_1'^2 t_2' + t_2'^2) - 
             q^3 (t_1^2 t_2 + t_2^2 + t_1^3 t_1'+\\&  
                t_1 t_1'^3 + t_2' (t_1'^2 + t_2'))) 
D_{t_2'})D_{t_2})D_{t_1}) \Biggr]\sum_\alpha \tau(t_1,t_2,\bar{t_1},\bar{t_2}, X_{\alpha}')\tau(t_1',t_2',\bar{t_1'},\bar{t_2'},X_{\alpha}^{\prime \prime})=0
\end{aligned}
\end{equation}

For the current setup \eqref{evol_op_U}, the necessity of use \eqref{split_exp} leads to the statement that such a way of the derivation of the bilinear identity on the $\tau$-function is not applicable for $U_q(sl_n)$ without any changes in the considered setting. To resolve it, one may refuse the property of commutativity of flows or replace the q-exponent from \eqref{evol_op_U} with the classical one.

\section{Towards $U_q(\mathfrak{sl}_n)$}

\subsection{Central element.}
It is well-known that the center of q-deformed UEA exists \cite{FRT} and its generators are given by.

\begin{equation}
    \mathcal{Z}_k=tr(q^{2\rho}(L^{(+)}S(L^{-}))^k), \quad k=1, \ldots, r=rank(\mathcal{G}),
\end{equation}

where $\rho=\frac{1}{2}\sum_{\alpha}H_\alpha$, $H_a$ is the basis of the Cartan subalgebra. 

We have the explicit formula for central elements $\mathcal{Z}_k$  of $U_q(\mathfrak{sl}_n)$ \cite{FRT}.
\begin{equation}
\tilde{\mathcal{Z}}_k=\sum_{\sigma, \sigma^{\prime} \in \operatorname{S_n}}(-q)^{l(\sigma)+l\left(\sigma^{\prime}\right)} L_{\sigma_1 \sigma_1'}^{(+)} \ldots L_{\sigma_k \sigma_k'}^{(+)} L_{\sigma_{k+1} \sigma_{k+1}'}^{(-)} \ldots L_{\sigma_n \sigma_n^{\prime}}^{(-)}, \quad k=1, \ldots, n-1 .
\end{equation}

Where L is an L-matrix \cite{Isaev}, and we take the sum over all permutations.
\[
L^+ = 
\begin{pmatrix}
q^{H_1} & 0 & \cdots & 0 \\
0 & q^{H_2} & \cdots & 0 \\
\vdots & \vdots & \ddots & \vdots \\
0 & 0 & \cdots & q^{H_N}
\end{pmatrix}
\begin{pmatrix}
1 & \lambda f_1 & \lambda f_{13} & \cdots & \ast \\
0 & 1 & \lambda f_2 & \cdots & \vdots \\
\vdots & \vdots & \ddots & \vdots & \vdots \\
0 & 0 & \cdots & 1 & \lambda f_{N-1} \\
0 & 0 & \cdots & 0 & 1
\end{pmatrix}
\]
\[
L^- = 
\begin{pmatrix}
1 & 0 & \cdots & 0 \\
-\lambda e_1 & 1 & \cdots & 0 \\
-\lambda e_{31} & -\lambda e_2 & \ddots & \vdots \\
\vdots & \vdots & \ddots & 0 \\
\ast & \cdots & -\lambda e_{N-1} & 1
\end{pmatrix}
\begin{pmatrix}
q^{-\tilde{H}_1} & 0 & \cdots & 0 \\
0 & q^{-\tilde{H}_2} & \cdots & 0 \\
\vdots & \vdots & \ddots & \vdots \\
0 & 0 & \cdots & q^{-\tilde{H}_N}
\end{pmatrix}
\]
It is possible to use the central elements to obtain BI on $\tau$-function for any specified rank of the algebra $U_q(sl_n)$ with the use of $\Delta Z_k \Delta X = \Delta X \Delta Z_k$.

\subsection{Fermionic realization} \label{ferm_real}
In this subsection, we use the free fermions formalism to rewrite the bilinear identity on $\tau$-function through q-intertwining operators. We note that there is a correspondence between the fundamental representations of $U_q(sl_n)$ and the free fermions algebra:
\begin{equation}
\left\{\psi_i, \psi_j^*\right\}=\delta_{i j} \quad\left\{\psi_i, \psi_j\right\}=0 \quad\left\{\psi_i^*, \psi_j^*\right\}=0
\end{equation}
The correspondence between the simple root generators of $U_q(sl_n)$ reads as
\begin{equation}\label{ferm_generators}
\begin{aligned}
e_i&=\psi_i \psi_{i+1}^* \\
f_i&=\psi_{i+1} \psi_i^* \\
h_i&=\psi_i \psi_i^*-\psi_{i+1} \psi_{i+1}^*
\end{aligned}
\end{equation}

While all other generators are given by
\begin{equation}
    e_{\alpha + \beta} = [e_\alpha, e_\beta]_{q^{\frac{1}{2}}},
\end{equation}
where q-commutator was defined in \eqref{qcom}. Now let us introduce the vector $|0\rangle$ such that
\begin{equation}
    \psi_{i}^{*}|0\rangle = 0 \quad \text{for } i \geq 1.
\end{equation}

Then, the highest weight vector in the $n$-th fundamental representation is given by
\begin{equation}
    |0\rangle_n = \psi_n\psi_{n-1} \ldots \psi_2\psi_1|0\rangle := |n\rangle
\end{equation}

and therefore the first fundamental representation consists of the $N$ vectors
\begin{equation}
    |1\rangle = \psi_1|0\rangle \quad \text{and} \quad   \mathbf{F}_{k-1}^{(1)}|1\rangle = \psi_k\psi_1^{*}|1\rangle = \psi_k|0\rangle, \quad \text{for } k = 2, \ldots, N
\end{equation}

or just
\begin{equation}
    \psi_k|0\rangle, \quad \text{for } k = 1, \ldots, N
\end{equation}

Similarly, the vector $\langle 0|$ in the same representation is defined as
\begin{equation}
    \langle 0|\psi_i = 0 \quad \text{for } i \geq 1 \qquad    \langle n| = \langle 0|\psi_1\psi_2 \ldots \psi_{n-1}\psi_n
\end{equation}

so that
\begin{equation}
    \langle n|\mathbf{E}_{k-1}^{(1)} = \langle n|\psi_1\psi_k^{*}, \quad \text{for } k = 2, \ldots, N
\end{equation}

Similarly, the second fundamental representation consists of the $N(N-1)/2$ vectors
\begin{equation}
    \psi_k\psi_l|0\rangle \quad \text{for } N \geq k > l \geq 1
\end{equation}

We choose the evolution operators in the same manner as before \eqref{evol_op_U}, while in terms of free fermions flows $\mathbf{E}_i, \mathbf{F}_i$ \eqref{E_bold} read as follows
\begin{equation}
\mathbf{E}_k:=\sum_i \psi_i \psi_{i+k}^*, \quad \mathbf{F}_k:=\sum_i \psi_{i+k} \psi_i^*
\end{equation}

These expressions correspond to the bosonization formulae. 

For simplicity in the current section, we perform the following change of variables the following one
\begin{subequations}\label{newcop}
\begin{align}
        &e_i\rightarrow e_i q^{-\frac{h_i}{2}}\\&
        f_i \rightarrow q^{\frac{h_i}{2}}f_i
\end{align}
\end{subequations}

Then the comultiplication reads, in new terms:

\begin{equation} \label{cm2}
\begin{aligned}
    \Delta e_i &= e_i \otimes K_i^{-2} + I \otimes e_i, \\
    \Delta f_i &= f_i \otimes I + K_i^{2} \otimes f_i,
\end{aligned}
\end{equation}

In \cite{KMM}, a q-deformed analogue of fermions as intertwining operators was presented, and we defined them as follows:
\begin{equation}
\Phi_i^{ R}=q^{2 \sum_{j=1}^{i-1} \psi_j^{} \psi_j^{*}} \psi_i^{ }, \quad \Phi_i^{ L}=q^{-2 \sum_{j=1}^{i-1} \psi_j^{} \psi_j^{*}} \psi_i^{},\quad, \Phi_i^{*,R}=q^{2 \sum_{j=1}^{i-1} \psi_j^{} \psi_j^{*}} \psi_i^{* }, \quad \Phi_i^{*, L}=q^{-2 \sum_{j=1}^{i-1} \psi_j^{} \psi_j^{*}} \psi_i^{*}
\end{equation}

where $\Phi^{ L}$,$\Phi^{*,L}$:$F_1 \otimes F_n \leftrightarrows F_{n+1} $ and $\Phi^{R},\Phi^{*,R}:F_n \otimes F_1 \leftrightarrows F_{n+1} $. These intertwining operators commute with the group-like element $g \in A(\mathcal{G})\otimes U_q(\mathcal{G})$ in the following sense
\begin{equation} \label{Gamma Uq}
    \Gamma = \sum_i \Phi_i^{L} \otimes \Phi_i^{*, R}, \qquad \Gamma\Delta g = \Delta g \Gamma.
\end{equation}

Moreover, it can be checked by explicit calculation that in the case of the non-group-like element $X \in U_q(sl_n)$, the property of commutation with the comultiplication \eqref{cm2} is preserved. This statement leads to the basic bilinear identity.
\begin{equation}
    \qquad \Gamma\Delta X = \Delta X \Gamma,
\end{equation}

however, as was mentioned in the section \ref{sec:fund_repr}, deriving bilinear identity on the $\tau$-function for $U_q(sl_n)$ with the evolution operators in the form of \eqref{evol_op_U} is possible only in the case of $n \leq 3$. Here we perform the derivation of BI for the fundamental representations of $U_q(sl_n)$. In the general form, BI reads:
\begin{equation}
\begin{aligned}
& \langle n| \exp_q \left(\sum_k t_k \mathbf{E}_k\right) \otimes\langle m| \exp_{q^{-1}} \left(\sum_k t_k^{\prime} \mathbf{E}_k\right) \Gamma \Delta(X) \exp_q \left(\sum_k \bar{t}_k \mathbf{F}_k\right)|n-1\rangle \otimes \exp_{q^{-1}} \left(\sum_k \bar{t}_k^{\prime} \mathbf{F}_k\right)|m+1\rangle= \\
& =\langle n| \exp_q \left(\sum_k t_k \mathbf{E}_k\right) \otimes\langle m| \exp_{q^{-1}} \left(\sum_k t_k^{\prime} \mathbf{E}_k\right) \Delta(X) \Gamma \exp_q \left(\sum_k \bar{t}_k \mathbf{F}_k\right)|n-1\rangle \otimes \exp_{q^{-1}} \left(\sum_k \bar{t}_k^{\prime} \mathbf{F}_k\right)|m+1\rangle
\end{aligned}
\end{equation}

We perform calculations similar to those done in the section \ref{Sec:BI}. To keep the notation from the non-deformed case, we can rewrite $\Gamma$ as follows
\begin{equation} \label{Gamma Uq int}
    \Gamma = \oint \frac{d z}{z} \Phi(z) \otimes \Phi^*(z), \qquad  \Phi (z) = \sum_{k=1}^{N} \Phi_k^{L} z^k, \qquad  \Phi^*(z) = \sum_{k=1}^{N} \Phi^{*,R}_k z^{-k}.
\end{equation}

After applying the q-BCH formula, we obtain the explicit expressions for the vertex operators for the first and second fundamental representations:
\begin{align} \label{vertex_op}
V_1(t_1,t_2,z) &= \frac{1}{(1+q^2)z^3}\Bigg[z\left((1+q^2)(q^2-zt_1)+(1-q^2)(zt_1^3+t_2+q^2t_2)D_{t_2}\right)\nonumber\\
&+\left(q^2+q^4(1-zt_1)-z^2(t_1^2-t_2)-q^2z(t_1+zt_2)D_{t_1}\right)\Bigg]\\
V_2(t_1,t_2,z) &= \frac{z}{q^2+q^4}\Bigg[z\left((1+q^2)(1+zt_1)+\left((1-q^2)(1+q^2+zt_1)t_1^2-(1-q^4)t_2\right)D_{t_2}\right)\nonumber\\
&-\left(q^4+z(t_1+zt_1^2+zt_2)+q^2(1+z(t_1+zt_2))D_{t_1}\right)\Bigg]\\
V_3(t_1,t_2,z) &= -\frac{z}{q^2+q^4}\Bigg[-\left(1+q^2\right)\left(z+q^2t_1\right)+D_{t_1}\left(\left(1+q^2\right)z^2+\left(1+q^2\right)zt_1+q^2t_2+q^4\left(t_1^2+t_2\right)\right)\nonumber\\
&+\left(-1+q^2\right)\left(zt_2+q^2\left(t_1^3+zt_2\right)\right)D_{t_2}\Bigg]\\
V_4(t_1,t_2,z) &= \frac{1}{\left(1+q^2\right)z^3}\Bigg[q^2\left(1+q^2\right)\left(z-t_1\right)+D_{t_1}\left(-\left(1+q^2\right)z^2+q^2\left(1+q^2\right)zt_1+q^2t_2+q^4\left(-t_1^2+t_2\right)\right)\nonumber\\
&+\left(-1+q^2\right)D_{t_2}\left(-q^2t_1^3+\left(1+q^2\right)z\left(t_1^2+t_2\right)\right)\Bigg]
\end{align}

In terms of the vertex operators, the bilinear identity has the form
\begin{equation}
\begin{aligned} \label{BI_vertex}
& \sum_\alpha \oint \frac{d z}{z} \left(\left[V_3\left(\bar{t}, z\right) \tau_n\left(t, \bar{t} ; X_\alpha^{\prime}\right)\right] \cdot\left[V_{4}\left(\bar{t}^{\prime}, z\right) \tau_m\left(t^{\prime}, \bar{t}^{\prime} ; X_\alpha^{\prime \prime}\right)\right]-\right. \\
&- {\left.\left[V_2(t, z) \tau_{n-1}\left(t, \bar{t} ; X_\alpha^{\prime}\right)\right] \cdot\left[V_{1}\left(-t^{\prime}, z\right) \tau_{m+1}\left(t^{\prime}, \bar{t}^{\prime} ; X_\alpha^{\prime \prime}\right)\right]\right)=0 },
\end{aligned}
\end{equation}

with chosen $n=2$, $m=1$. The obtained q-deformed vertex operators correspond to their classical limit 
\begin{equation} \label{vertex_cl}
V(t, z)=z^n \exp \left(\sum_k t_k z^k\right) \exp \left(-\sum_k \frac{z^{-k}}{k} \frac{\partial}{\partial t_k}\right).
\end{equation}

This algorithm may be generalized to $U_q(sl_n)$ after changing the evolution operators to ones that are q-commutative.

\subsection{R-matrix approach}
As a more complicated example of $\tilde{\Gamma}$ in $U_q(sl_n)$ that possesses the property $\tilde{\Gamma} \Delta X = \Delta X \tilde{\Gamma}$, let us consider the operator $\tilde{\Gamma} = R^{op} R$, where $R$ is a universal R-matrix. It can be realized as follows
\begin{equation}
    R^{op} R \Delta X = \Delta X R^{op} R,
\end{equation}
where the expression of $\mathcal{R}$-matrix was given in \eqref{Rmat}. We do not perform the calculations here because this approach is more complicated in comparison with the intertwining operators approach. All methods for deriving the BI result in identities that transform into each other with the redefinition of time parameters.

\section{Choosing evolution operators to be non q-deformed exponents.}

As mentioned above, the flows cannot be commutative for the choice of evolution operators in the case of $sl_n$ as in \eqref{evol_op_U}. Therefore one may refuse to use the q-exponential evolution operators \eqref{evol_op_U} and instead consider the following pair of operators.

\begin{equation} \label{evol_op_U_2}
    U(t)=\exp \left(\sum_{k=1}^{N-1} t_k \mathbf{E}_k^{(n)}\right), \quad
    \bar U(\bar t)=\exp \left(-\sum_{k=1}^{N-1} \bar{t}_k \mathbf{F}_k^{(n)}\right).
\end{equation}

Then there is no necessity for flows to be q-commutative, and we can save the commutativity of flows in the case of $sl(n)$. Here, we use the fermionic realization of $U_q(sl_n)$ introduced in the \ref{ferm_real} to illustrate the approach for obtaining BI on the example of $U_q(sl_3)$. The derivation repeats all steps from section \ref{ferm_real} and the changes appear only in the vertex operators \eqref{vertex_op}, which now have the form.

\begin{subequations} \label{vertex_exp_cl}
\begin{align} 
    &V_1=\frac{2 z (q^2 - z t_1) - (-1 + q^2) z (t_1^2 + 2 t_2)D_{t_2} + (2 q^2 + z (-2 q^2 t_1 + z t_1^2 - 2 z t_2)) D_{t_1}}{2 z^3}\\&
    V_2=\frac{z \left( 2 z (1 + z t_1) - (-1 + q^2) z  (t_1^2 - 2 t_2)D_{t_2} -  \left( 2 q^2 + z \left( t_1 (2 + z t_1) + 2 z t_2 \right) \right)D_{t_1} \right)}{2 q^2}\\&
    V_3=\frac{z \left( 2 z  t_1D_{t_1} + 2 (z + q^2 t_1) + (-1 + q^2) z  (t_1^2 - 2 t_2)D_{t_2} +  \left( 2 z^2 + q^2 (t_1^2 + 2 t_2) \right)D_{t_1} \right)}{2 q^2}\\&
    V_4=\frac{2 q^2 (z + t_1) + (-1 + q^2) z (t_1^2 + 2 t_2) D_{t_2} +  \left( -2 z^2 + q^2 ((-1 + 2 z) t_1 + 2 t_2) D_{t_1}\right)}{2 z^3}
\end{align}
\end{subequations}
These operators similarly  have the expected limit \eqref{vertex_cl}. The bilinear identity on $\tau$-function has the form \eqref{BI_vertex}, where we assume that the notation for the vertex operators \eqref{vertex_exp_cl} match with the previous section.

\section{Conclusion}

In this work, we have analyzed the non-group-like element approach \cite{2312, Bourg} for obtaining a bilinear identity on the $\tau$-function in terms of $U_q(sl_n)$. This approach allows one to get the bilinear identities for q-deformed UEA without appealing to the non-commutative algebra of functions, which is necessary when $\tau$-functions are made from the group-like elements $\Delta(g)=g\otimes g$. We explore various examples of such derivations of the bilinear identity on $\tau$-function, which can be based on different basic bilinear identities. Firstly, we derive the BI for the fundamental representation of $U_q(sl_3)$ with the use of the split Casimir approach. Instead of split Casimir, any element of the center of q-UEA may be used, however, the restriction on the fundamental representation is crucial for deriving the BI. The open question is whether the BI can be derived without such a restriction.

Also, we suggest a new approach to deriving the bilinear identity in the case of $U_q(\mathfrak{sl}_n)$ non-group-like element through the operator $\Gamma$. Although we possess convenient basic bilinear relations for any rank of q-deformed algebra, the relation on $\tau$-function could be derived only for $n\leq3$ because of badly defined evolution operators, which still lack an appropriate method for deformation. To clarify, the problem appears if one tries to restore the $\tau$-function in the matrix element of the basic bilinear identity. On the one hand, wishing to save the property of integrability in the classical sense, we use the set
of commutative flows in the system. On the other hand, the q-exponential, which usually appears during q-deformation,
cannot be differentiated as non-deformed and requires the q-commutativity of flows. Violation of this property is the main reason for the impossibility of obtaining a bilinear identity as an equation in the difference form. This argument leads to the indestructible statement that in the q-deformed case, the flows cannot commute with each other for a general rank of algebra if we preserve the q-exponential in evolution operators at the same time.

We suggest two possible ways of resolving the problem. Firstly, one may construct the flows that exhibit the property of q-commutativity and are devoid of commutativity. Secondly, differently defined evolution operators may be considered, where q-exponentials are replaced by common ones meanwhile there is no strong argument in favor of q-exponential in the evolution operators. In this paper, we studied the second way and presented the example of deriving BI based on $U_q(sl_3)$.

\section*{Acknowledgements}
The authors are greatful to A.Mironov, S.Mironov, A.Morozov, A.Popoliov, I.Ryzhkov for valuable discussions. The work was funded within the state assignment of the NRC "Kurchatov Institute".

\end{document}